# Delineating corneal elastic anisotropy in a porcine model using non-contact optical coherence elastography and ex vivo mechanical tests

Mitchell A. Kirby[1], John J. Pitre Jr. [1], Hong-Cin Liou[1], David S. Li[1], Ruikang Wang[1,2], Ivan Pelivanov[1], Matthew O'Donnell[1], Tueng T. Shen[1,2,*]

*[1]Department of Bioengineering, University of Washington, Seattle, Washington 98105, USA*
*[2]Department of Ophthalmology, University of Washington, Seattle, Washington 98104, USA*

* *Corresponding Author*

**Financial Support**

This work was supported, in part, by NIH grants R01-EY026532, R01-EY024158, R01-EB016034, R01-CA170734, R01-AR077560 and R01-HL093140, Life Sciences Discovery Fund 3292512, the Coulter Translational Research Partnership Program, an unrestricted grant from the Research to Prevent Blindness, Inc., New York, New York, and the Department of Bioengineering at the University of Washington. M. Kirby was supported by an NSF graduate fellowship (No. DGE-1256082). This material was based upon work supported by the National Science Foundation Graduate Research Fellowship Program under Grant No. DGE-1256082.

**Conflict of Interest**

No conflicting relationship exists for any author.

**Acknowledgements**

The authors wish to thank Dr. Yak-Nam Wang and the Center for Industrial and Medical Ultrasound at the University of Washington for their assistance in acquiring tissue samples. We also acknowledge Kit Hendrickson for preparing figure illustrations.

**Data Availability**

The authors declare that all data from this study are available within the Article and its Supplementary Information. Raw data for the individual measurements are available upon reasonable request.

**Running head**

Corneal elastic anisotropy: non-contact OCE and mechanical tests

**Abstract**

**Objective**: To compare non-contact acoustic micro-tapping optical coherence elastography (AμT-OCE) with destructive mechanical tests to confirm corneal elastic anisotropy.

**Design**: Ex vivo, laboratory study with non-contact AμT-OCE followed by mechanical rheometry and extensometry.

**Subjects:** Inflated cornea of whole-globe porcine eyes ($n$=9).

**Methods:** A non-contact AμT transducer was used to launch propagating mechanical waves in the cornea that were imaged with phase-sensitive OCT at physiologically relevant controlled pressures. Reconstruction of both Young's modulus ($E$) and out-of-plane shear modulus ($G$) in the cornea from experimental data was performed using a model of a nearly incompressible transversally isotropic (NITI) medium. Corneal samples were then excised and parallel plate rheometry was performed to measure shear modulus $G$. Corneal samples were then finally subjected to strip extensometry to measure the Young's modulus.

**Main Outcome Measures**: Strong corneal anisotropy was confirmed with both AμT-OCE and mechanical tests, with the Young's ($E$) and shear ($G$) moduli differing by over an order of magnitude. These results show that AμT-OCE can quantify both moduli simultaneously with a non-contact, non-invasive, clinically translatable technique.

**Results**: Mean of the OCE measured moduli were $E$=12.35 $\pm$ 5.23 MPa and $G$=30.90 $\pm$ 11.31 kPa at 5mmHg and $E$=20.05 $\pm$ 9.16 MPa $G$=60.61 $\pm$ 29.36 kPa at 20mmHg. Tensile testing yielded a mean Young's modulus of 1.39 MPa - 19.51 MPa over a strain range of 1-7%. Shear storage and loss modulus ($G'/G''$) measured with rheometry was approximately 81.79/12.76 $\pm$ 12.15/3.62 kPa at 0.2 Hz and 118.53 $\pm$ 10.39 kPa at 8.2 Hz (0.1% strain).

**Conclusions**: The cornea is confirmed to be a strongly anisotropic elastic material that cannot be characterized with a single elastic modulus. The NITI model is the simplest one that accounts for cornea's incompressibility and in-plane distribution of lamellae. AμT-OCE has been shown to be the only reported non-contact, non-invasive method to measure both elastic moduli. Sub-mm spatial resolution and near real-time operation can be achieved. Quantifying corneal elasticity in vivo will enable significant innovation in ophthalmology, helping to develop personalized biomechanical models of the eye that can predict response to ophthalmic interventions.

The cornea is one of the primary determinants of visual performance. Its structure of collagen fibrils embedded in a hydrated proteoglycan matrix forms a clear refractive component of the eye.[1,2] If corneal shape is not optimal, then images formed on the retina are aberrated. There are many methods to assess corneal shape, but there are no clinical tools to predict shape changes from interventions such as LASIK and collagen cross-linking therapies. Despite the overall success of these interventions over the last decades, outcomes remain unpredictable for an individual patient and many procedures produce unexpected changes in visual acuity and can have additional side effects. To optimize outcomes, a personalized biomechanical model of the cornea based on quantitative maps of mechanical moduli, and intraocular pressure (IOP) induced changes in mechanical moduli, is needed to predict final corneal shape.

Unfortunately, personalized biomechanical models have not been completely developed. Initial attempts have used technologies based on tonometry (e.g. Ocular Response Analyzer (ORA), Dynamic Scheimpflug Analyzer (DSA)) to estimate in vivo corneal mechanical properties as part of IOP measurements. Early results suggest that tonometry may be a screening tool for disease progression in common conditions related to elasticity, such as glaucoma and myopia.[3,4] However, it cannot determine fundamental material parameters required for robust biomechanical models of corneal deformation. In particular, tonometry metrics depend on experimental conditions and often characterize deformation in response to a dynamic mechanical stimulus over a large region of the cornea and sclera. In addition, tonometry does not consider the highly non-linear stress-strain relation between corneal tissue and pre-load (IOP), nor account for material anisotropy or variations in corneal thickness. Thus, to date, there are no non-invasive tools that can map corneal elasticity (accounting for its strong anisotropy) to provide the information needed for a personalized biomechanical model suitable for screening, surgical planning, and treatment monitoring.

Advances in the imaging speed and sensitivity of optical coherence tomography (OCT) helped launch OCT elastography (OCE), a technique to measure elastic moduli in the cornea at various length scales.[5] Recent developments in dynamic OCE produced technologies for non-contact excitation of propagating mechanical waves in the cornea and appropriate mechanical models for moduli reconstruction. Indeed, acoustic micro-tapping (AμT), which uses air-coupled ultrasound, is very effective at launching broadband, sub-mm wavelength mechanical waves in OCE. In addition, a nearly incompressible transversally isotropic (NITI) medium model describing corneal anisotropy has been used to estimate both (in-plane and out-of-plane) shear moduli in the cornea by tracking propagating wavefields.[6]

In our recent AμT-OCE ex vivo whole-globe study, we have shown that the cornea is a highly anisotropic elastic material, with up to three orders of magnitude difference between the in-plane Young's modulus (mainly driving its deformation and shaping) and the out-of-plane modulus (mainly responsible for out-of-plane shearing).[6] However, direct comparison between invasive mechanical tests and non-contact AμT-OCE, a necessary step in validating the method and helping translate it into a clinical tool, was not performed.

In this paper, we report non-contact AμT-OCE measurements of both elastic moduli in nine inflated porcine corneas of freshly excised whole eye globes at physiologically relevant controlled pressures and directly compare these measurements with mechanical tests performed on the same corneas immediately after AμT-OCE measurements. Parallel plate rheometry was used to measure cornea's out-of-plane shear modulus; whereas, to quantify the cornea's Young's modulus, tensile measurements were performed with an extensometer.

## Materials and Methods

### Porcine Cornea Samples

Porcine eyes were carefully enucleated from healthy adult animals (3-5 months old, 36-75 kg) by a specialist immediately following euthanasia. All excess tissue was removed to expose the sclera and the remaining whole globe was rinsed with an ocular hydrating solution (balanced saline solution, BSS). Whole globes were placed in a mold containing a damp sterile cotton pad to stabilize samples and mimic in vivo boundary conditions. A 20-guage needle connected to a bath filled with BSS was inserted through the temporal wall of the sclera to apply a controlled internal hydrostatic pressure (intraocular pressure, IOP). The IOP was controlled by raising and lowering the bath and was calibrated previously using porcine whole globes where an additional 20-guage needle was inserted into the wall of the sclera and attached to a digital hydrostatic pressure sensor. Following calibration, only a single needle insertion was performed in all AμT-OCE experiments. Each sample was held at the corresponding pressure for 5 minutes prior to scanning, over which a single drop of BSS was applied to prevent corneal dehydration. Each sample was scanned at room temperature and imaging took no longer than 1 hour per sample.

Following AμT-OCE, samples were removed from the imaging system. An incision was made approximately 2 mm from the limbus around the sclera to remove the cornea from the globe. The lens, iris, and ciliary body were carefully removed using tweezers. Corneal buttons were rinsed with BSS and then placed in a damped cloth and immediately transported for rheometry at room temperature. Rheometry measurements were repeated twice and took approximately 1 hour per sample. A drop of BSS was applied immediately following the scan to prevent dehydration prior to extension testing.

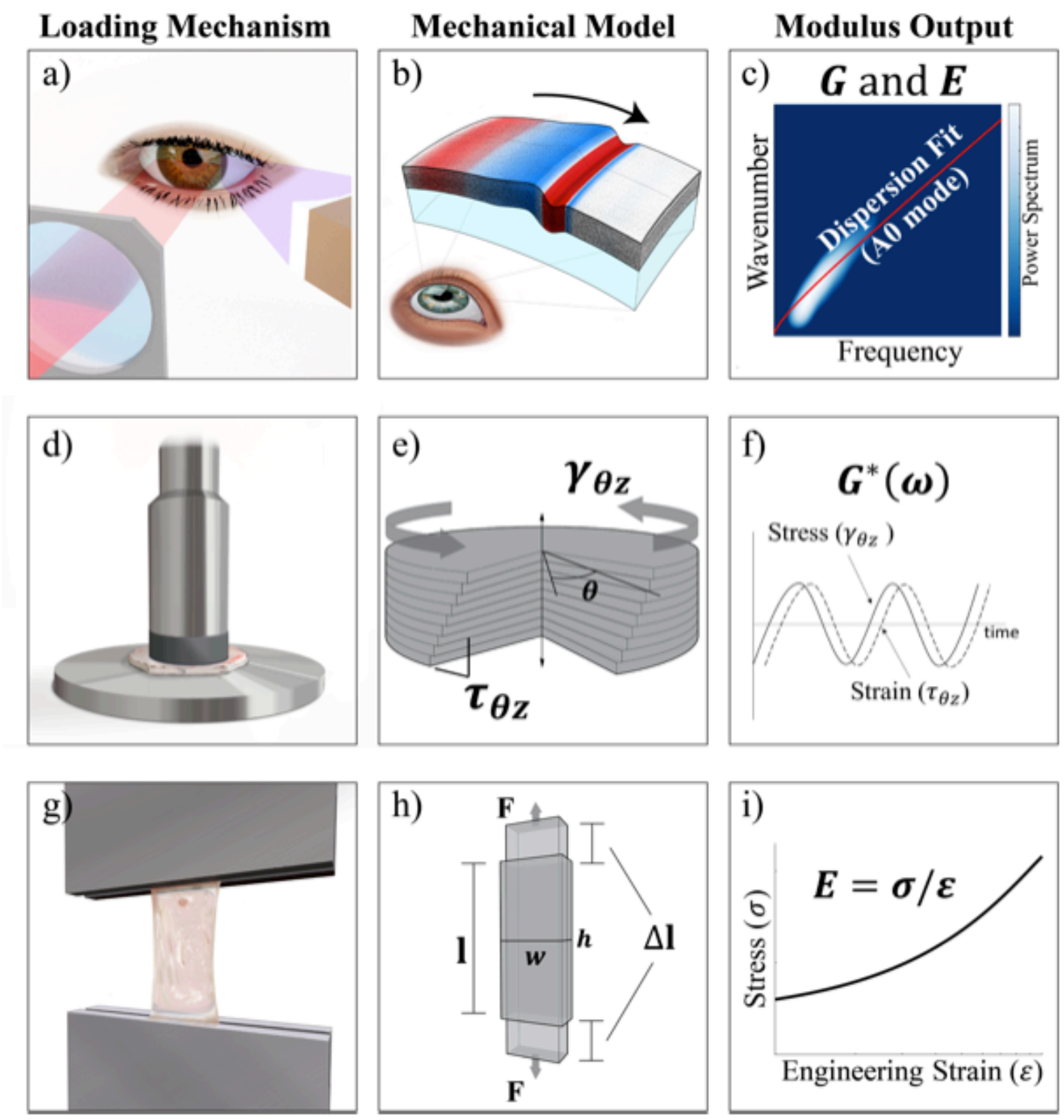

**Figure 1**. a) Representative display of AμT-OCE system to generate and image elastic waves in the cornea. The purple triangle focus represents non-contact acoustic loading through air and the red line-scan displays approximate OCT scan range. b) Guided elastic waves propagating in a NITI material such as the cornea. c) A solution to the dispersion equation was iteratively solved in the Fourier spectrum to reconstruct both elastic moduli, *E* and *G*, that most closely resemble the measured waveform in the cornea. d) Representation of ex vivo parallel plate rheometry of the cornea. e) Material model demonstrating assumed geometries and shear stress-strain relationships. f)

Calculation of the complex shear modulus $G^*$ using measured shear stress and strain. g) Representation of corneal strips loaded in extensometer. h) Material model demonstrating assumed geometries and loading in tensile testing. i) Calculation of the Young's modulus, $E$, in extension testing.

Cornea buttons were then each sectioned into strips approximately 6 mm wide along the nasal-temporal direction, leaving approximately 2mm of sclera on each end. The strips were transported for tensile testing in the same damp cloth and tested at room temperature within 1 hour of rheometry. Tensile loading and unloading was repeated 4 times. All data was acquired within 6 hours of animal euthanasia.

**Acoustic Micro-tapping Optical Coherence Elastography**

Inflated eyes were placed on a vertical translation stage and raised to the focal plane of the polarization-maintaining, phase-stable AμT-OCE system (detail provided in **Supplemental Methods**). A single AμT-OCE scan generated and tracked elastic waves using OCT operating in M-B mode. At the start of each M-B sequence, a trigger signal was sent to an air-coupled acoustic transducer providing a temporally and spatially focused acoustic 'push' pulse to create a localized displacement (tens-of-nanometer in amplitude) at the cornea's surface (**Figure 1a**) and launch a pulsed elastic wave propagating along the cornea's surface (**Figure 1b**).

Propagating elastic waves were tracked with OCT along the anterior corneal surface using a sequence of 512 repeated OCT A-scans (referred to as an M-scan); this sequence was conducted at 256 spatial locations in the nasal-temporal direction to form a 1.5mm × 10mm (axial × lateral) corneal scan for every time instant. The entire spatio-temporal scan took 1.33 seconds.

The AμT-OCE system was first tested using isotropic phantoms to validate the performance using a simple mechanical model (detailed in **Supplemental Methods**). Cornea whole globes were then scanned, and a two-dimensional Fourier transform applied to measured wavefields. A solution to the dispersion relation was found by varying the in-plane tensile and out-of-plane shear moduli to converge on a best-fit solution using a minimization routine applied to two-dimensional spectra (frequency-wavenumber (*f*-*k*) domain, detailed in **Supplemental**

**Methods**). This approach iteratively converged on a quantitative estimate of both the in-plane ($E$) and out-of-plane ($G$) shear moduli (**Figure 1c**). The corneal thickness ($h$) was determined using automated segmentation of the OCT image and served as a constraint on solutions to the dispersion relation.

**Parallel Plate Rheometry**

A parallel compression plate matching the approximate size of the cornea (12 mm diameter) clamped the anterior and posterior surfaces of the corneal buttons using a pneumatic rheometer (Anton Paar MCR 301 Physica) (**Figure 1d**). A relative twisting force (sinusoidal oscillatory shear stress) about an axis perpendicular to the surface was applied under a 5 N compressive pre-load and the resulting shear strain was measured to calculate the out-of-plane shear moduli, $G$, assuming a homogenous cylindrical material (**Figure 1e, f** corresponding equations in **Methods** and additional detail provided in **Supplemental Methods**).

**Stress-strain Extensometry**

Each cornea was cut into a strip and pneumatically clamped (2752-005 BioPuls submersible pneumatic grips, 250N max load) at the corneal-scleral boundary (**Figure 1g**). A 50 mN axial pre-load was applied and the samples were stretched at 2 mm/s (Instron model 5543) up to 10% strain. For each sample, two load-unload cycles were performed to precondition the tissue, followed by 3 rounds of force-elongation followed by relaxation. The Young's Modulus was determined assuming a beam-like structure (**Figure 1h**) and defined by the tangential slope of the stress-strain curve (**Figure1i**, corresponding equations in **Methods** and additional detail provided in **Supplemental Methods**).

## Results

**Determination of both Corneal Elastic Moduli with AμT-OCE**

The cornea was modeled as a layer of a nearly incompressible transversally isotropic (NITI) material bounded by air on top and by water on bottom. As shown previously, the cornea can be treated as a flat NITI layer for wave propagation because its thickness is much smaller than its radius of curvature.[6] Consequently, guided waves propagating in the cornea exhibit a unique dispersion relationship defined in the NITI model by the Young's modulus ($E = 3\mu$) and transverse shear modulus ($G$) (detailed in **Supplemental Methods**). The stress response to an induced strain for such a material is described using Hooke's law:

$$\begin{bmatrix}\sigma_{xx}\\ \sigma_{yy}\\ \sigma_{zz}\\ \tau_{yz}\\ \tau_{xz}\\ \tau_{xy}\end{bmatrix} = \begin{bmatrix}\lambda+2\mu & \lambda & \lambda & & & \\ \lambda & \lambda+2\mu & \lambda & & & \\ \lambda & \lambda & \lambda+2\mu & & & \\ & & & G & & \\ & & & & G & \\ & & & & & \mu\end{bmatrix}\begin{bmatrix}\varepsilon_{xx}\\ \varepsilon_{yy}\\ \varepsilon_{zz}\\ \gamma_{yz}\\ \gamma_{xz}\\ \gamma_{xy}\end{bmatrix}. \tag{1}$$

where $\sigma_{ij}$ denotes engineering stress, $\varepsilon_{ij}$ denotes engineering strain, $\tau_{ij}$ denotes shear stresses, $\gamma_{ij} = 2\varepsilon_{ij}$ denotes shear strains, and the subscripts $x$, $y$, and $z$ refer to standard Cartesian axes (a detailed description of elastic waves in NITI materials can be found in Pitre et. al.[6]). Note that in nearly incompressible media (such as all soft biological tissues) the longitudinal modulus $\lambda$ does not influence medium deformation although it is much larger than $\mu$ and $G$, and the Young's modulus $E$ is simply $3\mu$. Thus, functional biomechanical models do not require estimating λ.

The AμT-OCE system (see **Methods** and **Supplemental Methods**) was used to track propagating mechanical waves in porcine corneas. In all scans, wave energy was concentrated in a single dispersive wave mode (consistent with the A0-mode) and provided a reliable fit with the dispersion relation for a NITI model. Measured wavefields were processed as described in Ref 6 to reconstruct both corneal moduli $E$ and $G$ (**Figure 1g**).

Elastic moduli were reconstructed from OCE data for nine ($n$=9) cornea. On average (statistical mean), in-plane Young's modulus was 12.35 $\pm$ 5.23 MPa at 5mmHg and 20.05 $\pm$ 9.16 MPa at 20 mmHg. The mean transverse shear modulus was 30.90 $\pm$ 11.31 kPa at 5mmHg and 60.61 $\pm$ 29.36 kPa at 20 mmHg. The inflation pressure placed the cornea in a state of pre-stress, which varied in magnitude according to the IOP. This result demonstrates that both in-plane tensile and out-of-plane shear moduli increase with increasing IOP, consistent with the non-linear material properties of the cornea. The cornea thinned on average by 50 μm as the pressure was increased. In all samples the endothelium appeared intact.

**Determination of the out-of-plane shear modulus with Parallel Plate Rheometry**

Corresponding corneal buttons were deformed in a state of pure transverse shear strain according to:

$$\gamma_{\theta z}(t) = \gamma_{\theta z_0} e^{i\omega t}\ , \tag{2}$$

where $i = \sqrt{-1}$, $\omega$ is the angular frequency, *t* is time, and $\gamma_{\theta z_0}$is the peak shear strain amplitude. Collagen-rich tissue generally is linearly elastic below 1%,[7] thus 0.1% peak shear strain was applied during the frequency sweep. The shear stress ($\tau_{\theta z}$) is described by:

$$\tau_{\theta z}(t) = \tau_{\theta z_0} e^{\left(i(\omega t+\delta)\right)} \tag{3}$$

where $\tau_{\theta z_0}$ is the shear stress amplitude, $\delta$ is the phase shift angle between applied stress and the corresponding shear strain. The shear stress and strain relations can be described by:

$$\tau_{\theta z}(t) = G^*(\omega)\gamma_{\theta z}(t) \tag{4}$$

where the complex shear modulus, $G^*(\omega) = G(\omega)' + iG(\omega)''$, is defined by the in-phase (storage ($G'$)) and out-of-phase (loss ($G''$)) stress-strain relations. Frequency-dependent storage and loss moduli of the complex *G* value in the NITI model (detailed in **Supplemental Methods**) were determined over a range of .16-16 Hz. Sample thickness was recorded based on the parallel plate gap distance.

The shear storage and loss modulus (*G*'/ *G''*) measured in corresponding porcine corneal-buttons subject to frequency-swept rotational shearing ranged from 81.79/12.76 $\pm$ 12.15/3.62 kPa to 132.6/28.91 $\pm$ 15.86/2.65 kPa for 0.16 - 16 Hz at 0.1% peak strain amplitude. Corneal buttons were tested using the smallest compressional pre-load that could be applied without tissue slippage and incorporated a portion of the scleral rim. The unconfined compression caused a portion of the tissue to splay beyond the custom-printed disk during testing. A single frequency sweep took approximately 30 minutes, over which each sample progressively thinned about 27µm.

**Determination of corneal Young's modulus with Stress-strain Extensometry**

The engineering stress and strain were calculated using the cross-sectional area of the cornea (measured with a digital micrometer) and force-elongation recordings assuming a beam structure (detailed in **Supplemental Methods**). In this configuration, the tissue was subject to uniaxial tension along the nasal-temporal ($xx$) direction and the corresponding stress-strain relation described by:

$$E = \frac{\sigma_{xx}}{\varepsilon_{xx}}, \qquad (5)$$

where the strain-dependent Young's modulus was defined as the instantaneous slope of the stress-strain curve. The tangential (Young's) modulus, $E$, measured in tensile testing over a range of engineering strain from 1-10% was 1.39 - 40 MPa (**Supplemental Methods**). The orientation of the strip lay in the $xy$-plane and the direction of the loading stress was uniaxial along the nasal-temporal axis. Corneal strips were on average 5.95 mm wide, resulting in a cross-sectional area of 4.96 $mm^2$.

It can be shown that for the $xx$-loading direction in a NITI material, $E$ is sensitive only to the in-plane elastic modulus (detailed in **Supplemental Methods**). Based on this analysis, AµT-OCE provides a functionally equivalent measure of the elastic moduli probed in both shear rheometry ($G$) and tensile extension ($E$) but in a single, non-contact acquisition that can be performed non-invasively using intact whole-globe eyes.

**Comparison Between Methods**

A direct comparison between the mean and standard deviation of the moduli measured via OCE, shear rheometry, and strip extensometry is shown in **Figure 2**. Most notably, both the tensile elastic modulus ($E$) and shear modulus ($G$) differ by approximately an order of magnitude, consistent with previous results.[6]

To compare OCE tests of inflated cornea with rheometry and extensometry, boundary conditions and equivalent strain effects were estimated. In rheometry, the scleral rim was retained to provide a surface to grip during tensile loading. To account for the effect of excess sclera, three separate cornea buttons were tested with the scleral ring and then immediately after removing the rim. The results (**Figure 2,** detailed in **Supplemental Methods**) indicate that the excess tissue induced an approximately 2-fold increase in the shear modulus compared to rheometry of the cornea alone. Consequently, a scaling factor was applied to rheometry results. Both the apparent (with sclera) and adjusted (without sclera) shear modulus range is displayed for comparison. Of note, the NITI model assumes a negligible viscosity. Thus, we compare the *G* value from OCE with the storage modulus (*G*') in rheometry.

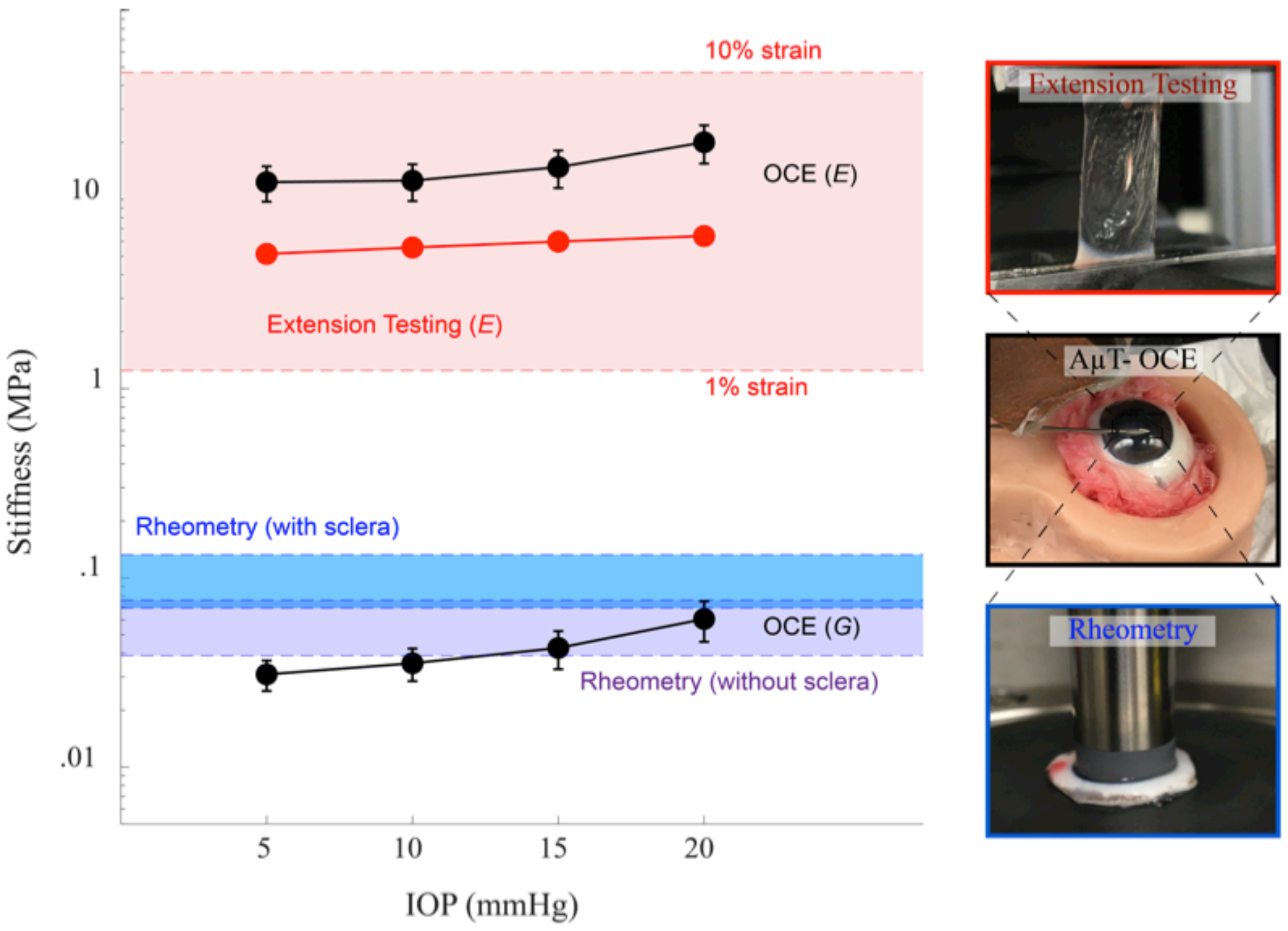

**Figure 2.** Moduli (*E* and *G*) estimates from AμT-OCE (black), parallel-plate rheometry (blue) and strip extension (red). The red box denotes the range of elastic moduli measured at strains of 1%-10% and the red dots denote the equivalent strain estimate at intraocular pressures representative of in vivo conditions (detailed in **Supplemental Methods**). The darker blue box is the storage shear modulus (*G'*) measured via rheometry with the scleral ring attached and the light blue box corresponds to values appropriately scaled to account for the apparent stiffness from inclusion of the scleral ring (detailed in **Supplemental Methods**).

As the cornea inflated via IOP is under a state of tension, an estimated strain offset was introduced to account for tensile strain introduced due to IOP as well as compressive forces within the anterior portion of the cornea introduced when flattening a curved structure (detailed in **Supplemental Methods**) during elongation. The strain offset first references the state where strips were exposed to only tension, then, the internal pressure (within the tissue) was used to predict equivalent strain values within the samples loaded via inflation and strip extension. At an IOP of 5-20 mmHg, the equivalent engineering tensile strain was estimated to be between 3.97% $\pm$ 0.17% and 4.44% $\pm$ 0.14 %. The Young's Modulus estimate from OCE corresponded to an equivalent strain between 5.72% and 7.10% in strip extensometry. The average Young's modulus measured via strip extension was 1.24-46.8 MPa for 1%-10% strain, respectively.

**Discussion**

Predicting corneal shape changes after medical interventions, and longitudinal alterations from post-procedure performance, is critically important to improve screening, surgical planning, treatment monitoring and overall outcomes in vision correction therapies. Due to the structural complexity and mechanical nonlinearity (i.e., IOP dependent moduli) of the cornea, a personalized mechanical model is required to optimize procedure outcomes. Such a model must include in vivo measurements of topography, IOP, and maps of corneal mechanical moduli. As the IOP is not consistent and the corneal elastic moduli are nonlinear, moduli changes induced by variations in IOP must also be considered. To-date, reliable measurements of ocular mechanical properties have only been possible on ex vivo samples and have not directly impacted the clinic.[8] While critical to our understanding of corneal mechanics, the destructive nature of most traditional techniques render these approaches impractical for clinical translation.

Anterior segment OCT can map corneal shape, and tonometry can estimate IOP. However, only recently a non-contact method has been shown to provide quantitative information on corneal elasticity. In our recent study, we showed that non-contact AμT-OCE can simultaneously assess both corneal moduli, *E* and *G*, under physiologic loading conditions.[6] It was very important that the AμT-OCE measured in-plane Young's modulus, *E,* was in good correspondence with literature results obtained by destructive ex vivo inflation and tensile tests; whereas the out-of-plane shear modulus *G,* being a few orders of magnitude smaller than *E,* reasonably matched literature data on the shear modulus obtained by rheometry.

In this study, we performed one-to-one comparison of corneal moduli measured with AμT-OCE with two mechanical tests (parallel plate rheometry and tensile extensometry) performed on the same corneal samples. Nine fresh porcine corneas were measured to limit individual variations in the cornea's mechanical properties. While this study demonstrates markedly different corneal stiffness under shear versus tensile loading, there are some differences between AμT-OCE and destructive mechanical testing methods that can make direct comparisons difficult. Indeed, effects such as corneal curvature, its nonlinearity, boundary conditions in mechanical loading, loading direction, engineering strain, hydration, pre-conditioning and others can influence moduli estimates in mechanical tests. Estimates provided by AμT-OCE should be more accurate representations of in vivo biomechanics because this non-contact and non-invasive measurement procedure was performed under well controlled IOP.

In extension loading, corneal shape is different from inflation. The cornea preserved its curvature in OCE measurements, while the curvature was flattened in tensile testing. Axial extension, which causes the cornea to flatten, induces compressional forces and produces different strain distributions for cornea under inflation versus extension. Additionally, the cornea was under biaxial tension in OCE and uniaxial tension in extension. The force is assumed equally distributed across the cornea considering a true cuboid shape (uniform thickness and cross-section). However, the central cornea thickness is thinner than the periphery and exhibits varied radius along the center line compared to the edge of the strip, complicating interpretation of results.[9] Corneal samples were not uniform thickness, suggesting that the axial strain at the center of the sample may differ from that at the perimeter during loading. This effect that was not adjusted for in this analysis. Nevertheless, our measurements of Young's modulus acquired from both AμT-OCE and extension testing are consistent with those in in porcine cornea strips subject to tensile loading (0.3-29 MPa[10–16]).

The order-of-magnitude difference in the modulus measured under shear loading was higher than previously reported values of $G$ (0.34-9 kPa[17–19]). Because corneal-buttons were loaded in unconfined compression, tissue at the corneal-scleral junction splayed outward beyond the probe. Rheometry was performed on an independent set of corneas both with and without the scleral rim attached (**Supplemental Methods**). The remaining tissue was shown to produce an approximately two-fold increase in the measured modulus. This suggests that the tissue splayed beyond the probe provides additional resistance to shearing, over estimating the corneal modulus. Additionally, the corneal-scleral boundary may contribute to residual stress within the cornea compared to the corneal button alone.[20] The shear-modulus of the cornea has been shown to increase with higher compressional pre-load[17], suggesting that internal forces due to both parallel-plate rheometry and increased residual stress may increase the measured transverse shear modulus. The hypothesis that residual stress within whole-globe samples may contribute to a higher transverse shear modulus likely warrants further study and may describe the discrepancy between the values measured via OCE and those reported in the literature. Internal pressures estimated in the cornea due to flattening during tonometry do not appear to induce large structural changes within the tissue at low IOP, suggesting that flattening alone will have a small effect on the shear modulus.[21] In this case, flattening due to the parallel-plate arrangement was assumed to have a small effect on the measured modulus. It should be noted however that during rheometry, tissue at the center undergoes a smaller shear strain than that at the outer edge. The center of the tissue may also experience different compressive strain due to corneal thinning, an affect that was ignored in this study. Additionally, the frequency range used by OCE differed from that of rheometry. While OCE functionally measures higher frequency vibrations (0.3 kHz – 4 kHz range), the relatively lower frequency (multiple Hz-range) probed in rheometry suggest that frequency-dependent differences likely exist between the moduli measured by the two methods. Finally, Young's and shear moduli depend on depth in the cornea, an effect not included in this study. Viscous effects also warrant additional exploration.

This study suggests that robust and accurate measurements of corneal elastic moduli can be achieved using non-contact AμT-OCE. While in-plane anisotropy (within the xy-plane) has been reported in some studies, the degree of anisotropy remains low at low IOP.[14,22–25] This suggests that for normal IOPs (below 25 mmHg in porcine cornea), corneal microstructure can be approximated with the NITI model (i.e., symmetric for any direction in the *xy*-plane). It should be noted that significant effort has been directed toward personalized biomechanical models suitable for screening, surgical planning, and treatment monitoring. Because corneal mechanical properties determine its shape, it is important that accurate moduli are input to biomechanical models for predicting the static deformation and shape.[26] While such static models remain largely in the development stage, advanced models assuming a transverse isotropic tissue structure appear most robust.[27,28]

Because AμT-OCE appears to accurately measure anisotropy of mechanical properties in the cornea, while OCT simultaneously can image its structure and shape, an OCT-based technique is a very promising direction to develop personalized biomechanical models. Such personalized models can potentially be used to study disease progression and may play a role in treatment based on simulated interventions. Future studies to assess anisotropy in diseased cornea may further improve our understanding of pathologies and potentially inform treatment. Additional studies using higher resolution OCE reconstruction methods may even increase the utility of such models. Assuming in vivo OCE methods can robustly and reliably measure moduli, it is likely that truly non-contact methods will pave the way for clinical translation.

## Supplementary Online Content

# Delineating corneal elastic anisotropy in a porcine model using non-contact optical coherence elastography and ex vivo mechanical tests

Mitchell A. Kirby[1], John J. Pitre Jr. [1], Hong-Cin Liou[1], David S. Li[1], Ruikang Wang[1,2], Ivan Pelivanov[1], Matthew O'Donnell[1], Tueng T. Shen[1,2,*]

*[1]Department of Bioengineering, University of Washington, Seattle, Washington 98105, USA*
*[2]Department of Ophthalmology, University of Washington, Seattle, Washington 98104, USA*

**Supplemental Methods.**

## Supplemental Methods

### 1. Acoustic micro-tapping Optical Coherence Elastography (AμT-OCE) system

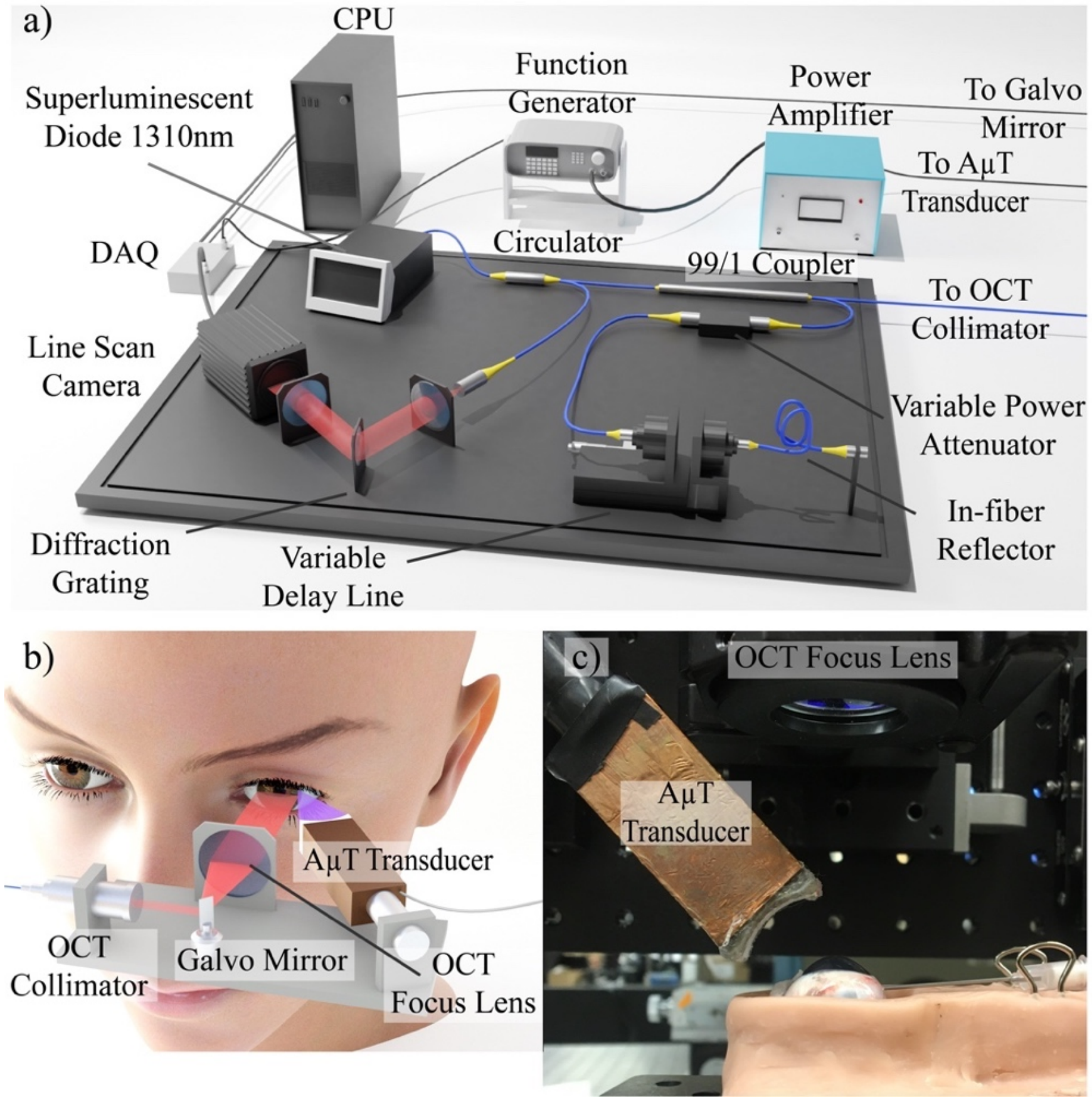

**Supplemental Figure 1**. a) AμT-OCE system components and connections based on a phase-stable polarization maintaining SD-OCT configuration. b) Close-up of AμT-OCE imaging arm showing optical and acoustic focus alignment on a stylize human for display. c) Photograph showing experimental set-up for testing whole globe porcine cornea inflated to physiologically relevant intraocular pressures.

A phase-sensitive spectral domain optical coherence tomography system (**Supplemental Figure 1**) was built to track propagating mechanical waves via local tissue displacements. The system utilized a Michelson-type fiber-optic interferometer combined with an optical spectrometer and line scan camera for depth encoding. The system was optimized for noise reduction using polarization-maintaining (PM) fibers in all components (as opposed to conventional single mode (SM) fibers). The light source was a broadband superluminescent diode (SLD) with a near-Gaussian spectral profile and maximum output power of 30mW (SLD1018P, Thorlabs, NJ) coupled into a PM-fiber with radiation aligned along the slow axis. The central wavelength was 1310 nm, with a FWHM of 45 nm. Linearly polarized light was then coupled into a broadband PM circulator (PMCIR-1310_FC/APC, OZ-optics, Ottawa, Canada) and then into a 2x2 99/1 PM beam combiner (Fused-22-1310-7/125-99/1, OZ-optics, Ottawa, Canada). One percent (1%) of the power was directed toward the reference arm through a variable attenuator (VOA50PM, Thorlabs, NJ) and into a variable optical delay line (ODL-200, OZ-Optics, Ottawa, CA) terminated by an in-fiber reflector having near 100% reflection at the 1310 nm wavelength. The NA of the imaging arm was 0.057, providing a lateral resolution of 14.72 μm. The optical spectrometer consisted of a collimator ($f = 50$ mm), transmission grating (1200 lines/mm), achromatic focusing lens ($f = 100$ mm), and a high speed 1024-pixel line scan InGaAs CCD camera (1024-LDH2, Sensors Unlimited, New Jersey, USA). Dispersion was numerically corrected, and the effective axial resolution was ~20 μm in air, as experimentally measured treating a mirror surface as a point-source.

To generate and track propagating shear waves, the OCT system operated in M-B mode, wherein a trigger signal was sent to a home-built air-coupled ultrasound transducer providing a temporally and spatially focused acoustic wave that transferred energy to the tissue, effectively generating a tens of nanometer to micron-scale displacement. At the beginning of a sequence of 512 repeated A-scans, (referred to as an M-scan), at each location a 600 V peak-to-peak, 100μs-long, linear chirp (bandwidth of 1 MHz to 1.1 MHz) was delivered to a piezoelectric transducer. More details on the transducer can be found in Supplemental Reference 1. The pulse/M-scan sequence was captured at 256 spatial locations, 54.7 μm between adjacent locations, across the image plane sequentially to create a 3-D cube (*z, x, t*). A complete M-B scan consisted of 1024 depth × 256 lateral locations × 512 temporal frames (captured at 90,000 frames per second) with an effective imaging range of 1.5 mm × 10 mm (axial × lateral). One full M-B scan took 1.33 s. Shear wave generation and OCT beam scanning was precisely controlled using custom LabVIEW software. More details on the excitation and detection routines can be found in Supplemental Reference 2.

The resulting three-dimensional dataset was then used to reconstruct the propagating wave based on the OCT-measured local particle vibration velocity. The axial vibration velocity at a given location ($v_z(x, z, t)$) was obtained from the optical phase difference between two consecutive A-line scans at each location. The optical phase stability of 8 mrad (50dB OCT-SNR) was determined using the standard deviation of the phase signal[3] at the surface of a static glass coverslip, indicating sensitivity to vibration amplitudes as small as 600pm between successive A-scans.

To analyze all surface vibrations, the signal along the curvilinear surface of the cornea was used after automatic identification with an edge detection algorithm. Phase data in a 183μm window below the surface were extracted and averaged using spatial weighting with one half of a Gaussian window (HWHM = 90 μm, weight decreasing with depth).

**2. Phantom Measurements: Comparison of AμT-OCE with mechanical Methods**

As there are no current 'gold-standard' techniques to perform non-contact mapping of moduli, isotropic phantom measurements were compared with destructive mechanical testing techniques. The elastic moduli of both thick and thin (relative to the shear-wave wavelength) mechanically isotropic phantoms without inflation pressures were determined using AμT-OCE and compared with results of shear rheometry and tensile testing.

Mechanically isotropic PVA-based elastic phantoms were prepared based on a previous study.[4] First a 4:1 ratio of dimethyl sulfoxide (DMSO, CAS: 67-68-5, EMD Millipore Corp.) was mixed into water using a stir plate. A concentrated stock solution of 0.3 wt. % titanium nanoparticles suspended in water was prepared using a tip sonicator (Digital Sonifier 450, Branson, Danbury, CT, USA) for 3 minutes (30% duty cycle) at an amplitude of 30% to disperse nanoparticles in solution and tune the phantom's optical properties. The nanoparticle solution was added to the DMSO solution to achieve a nanoparticle concentration of 0.025 wt. %. Two (2) wt. % PVA (146-186 kDa, >99% hydrolyzed, CAS: 9002-89-5, Sigma-Aldrich) was then added to the solution. It was covered and stirred on a hot plate maintaining a temperature of 120° C for approximately 1 hour to dissolve the PVA. Once fully dissolved, the solution was degassed using a vacuum chamber to remove any bubbles.

Molten PVA solution was then poured into circular molds with a diameter of 5 cm to cast multiple phantoms with variable thickness. Thin slabs were created by adding just enough molten PVA solution to cover the bottom of the mold and then rotating the mold to evenly distribute the solution before storing at -20°C for up to 12 hours to solidify. Thicker phantoms were created by pouring molten PVA into the mold up to heights of 4cm.

Casted phantoms were allowed to harden at room temperature and then placed in a water bath to diffuse DMSO out of solution. The bath was changed regularly for a minimum of 48 hours to remove all DMSO and the phantoms were stored in deionized water to prevent dehydration. The thin PVA phantom was suspended on top of water to force asymmetric boundary conditions and measured using OCT structural imaging. The thickness in the thin phantoms ranged from 400μm-1mm, as measured with OCT and verified with a digital micrometer.

Elastic moduli were quantified using OCE measurements performed on both bulk and layered samples from the same phantom batch. In the layered sample, the Young's modulus ($E$) was quantified by fitting the guided modes in FK-space to the A0-mode of the isotropic thin-plate solution.[4–6] The modulus in the thick phantom was quantified using the group velocity of the Rayleigh wave ($C_R$) and directly inverting to obtain the Young's modulus according to

$$E = 3\rho \left( \frac{C_R}{0.955} \right)^2, \tag{S1}$$

where $\rho$ is the density (assumed to be $\rho = 1000 \frac{\mathrm{kg}}{\mathrm{m}^2}$). A Radon Sum method[7] was used in time-of-flight reconstruction of the surface wave group velocity in thick phantoms. As demonstrated previously, the difference in moduli quantified from thick and thin phantoms was less than 3%.[4,6] The moduli from a total of n=5 thin (~1mm) and n=5 thick (~4cm) phantoms were averaged to get the OCE-measured elastic modulus (**Supplemental Figure 2a**).

Following OCE measurements, circular slabs were cut from all phantoms and tested with a parallel plate rheometer (Anton Paar MCR 301 Physica). The frequency-dependent out-of-plane shear behavior was measured over a range of 0.1-16 Hz using a 25 N compressive preload and a peak shear strain of 0.1%. This yielded estimates of the frequency-dependent storage and loss moduli, corresponding to the real and imaginary parts of the complex shear modulus (**Supplemental Figure 2a**). In a lossless isotropic media, the shear storage modulus is proportional to the Young's modulus according to $E = 3\mu$.

The Young's modulus was directly measured in stress-strain testing using linear strips (~6 mm wide) cut from samples and pneumatically clamped (2752-005 BioPuls submersible pneumatic grips, 250N max load) at a width of ~15mm. A 50mN pre-load was applied. The samples were stretched at 2mm/s (Instron model 5543) over a range of 1-10% strain and repeated across 3 cycles. The Young's modulus was then computed from the instantaneous slope of the stress-strain curve over this strain range (**Supplemental Figure 2b**).

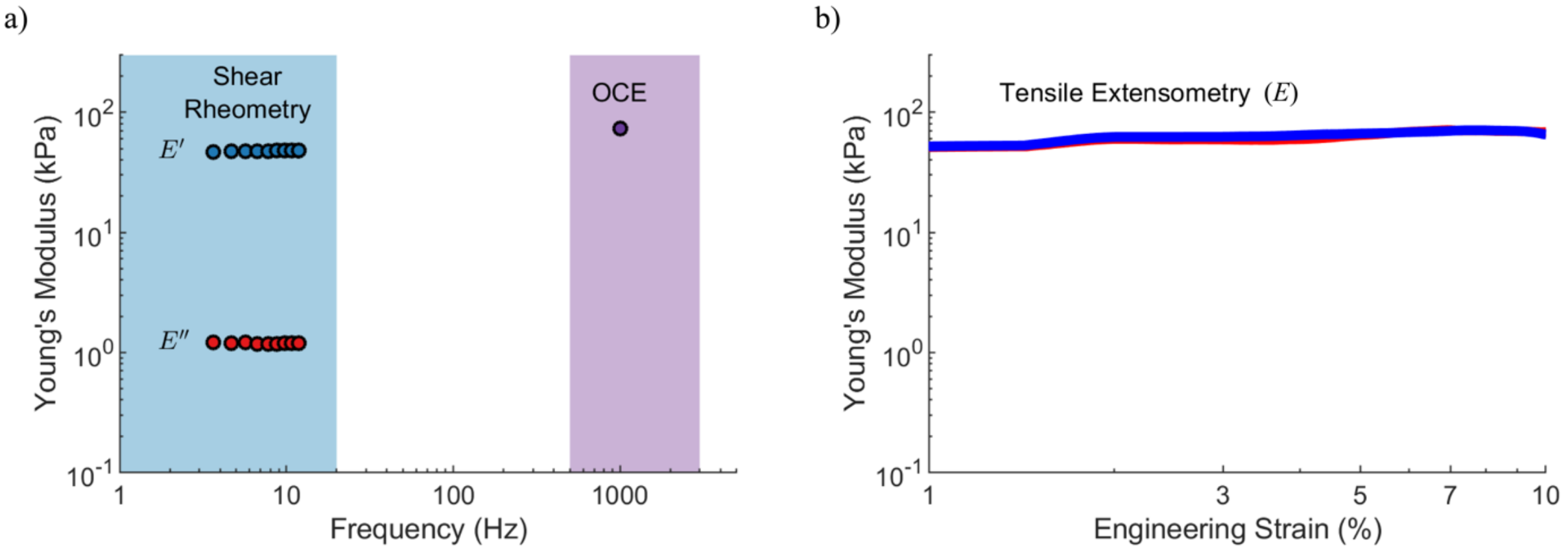

**Supplemental Figure 2.** Comparison between Young's modulus in a PVA phantom measured with a) parallel-plate rheometry and AμT-OCT. b) The mean tangential modulus for multiple strips taken from the thin (red) and thick (blue) phantoms.

The Young's modulus quantified in extensometry, AμT-OCE, and rheometry was 64.76 kPa, 73.79 kPa, and 47.67 kPa, respectively. The lossless isotropic material model was supported by the frequency-dependent rheometry and strain-dependent tensile test results. The storage (*E*') modulus from rheometry is close to, but slightly lower than, the Young's modulus from OCE based on the average of measurements for multiple (n=10) tissue-mimicking (PVA cryogel) phantoms. For all measurements, the ratio between OCE and rheometry was in the range of 1.3-1.5 (30%-50% error). These results are consistent with Minton et.al[8], where a significant poro-/viscoelastic effect in PVA phantoms was shown to produce an apparent stiffness (1.57-1.71 times lower than the peak stiffness) under sustained compression in rheometry. Note that unlike phantom materials such as agarose with high viscosity[9], PVA's low viscosity should have little effect over the frequency range of our OCE measurements (0.5 kHz – 3 kHz). On average, the ratio between OCE and extensometry was 1.14 (14% error). Even with the poro-/viscoelastic effect of compression in PVA material, OCE-measured moduli was within 50% of the values determined by destructive testing in both methods. This confirmed that when analyzed with the appropriate mechanical model, AμT-OCE can accurately determine equivalent elastic moduli compared to destructive testing techniques. Because AμT-OCE was shown to reliably quantify elastic moduli without damaging soft material, we conclude that the system may reliably be used to quantify moduli in the cornea using the NITI mechanical model.

### 3. Corneal Moduli quantification in AμT-OCE

Quantitative moduli estimates can be found that most closely match guided wave modes in the cornea using a fitting routine based on the full frequency-wavenumber (FK) spectra of the OCE-measured surface vibration signal.[10] While the solution includes an infinite number of quasi-symmetric and quasi-antisymmetric modes, AμT-generated wave energy concentrated in the A0 dispersive wave mode. Thus, only the A0 mode was fit to the NITI dispersion relation. Using the dispersion curve solution as the forward model, a mode-tracing routine was applied to fit the theoretical dispersion relation to the measured data by maximizing the following objective function:

$$\Phi(\mu, G) = \frac{1}{N_f}\sum_f \sum_k w(f, k; \mu, G)|\hat{v}(f, k)|^2 - \mathrm{B}\left|\frac{\mu}{\lambda}\right| \quad \text{(S2)}$$

where $\hat{v}$ is the normalized 2D Fourier spectrum of the measured surface velocity, $w(f, k; \mu, G)$ describes the $A_0$ mode solution, $N_f$ is the number of FFT bins included in the fit, and B is a regularization term that ensures the ratio between $\mu/\lambda$ remains small (satisfying the nearly-incompressible assumption). B was set to 1, based on an L-curve analysis.[11] The value of $\lambda$ was updated at each iteration and for a given frequency $f$ and given the current parameter set $(\mu, G)$, the dispersion relation solver returned the wavenumber $k_0$ associated with the $A_0$ mode. A weighting function is applied to each frequency with a peak value at the wavenumber of the theoretical dispersion curve, with weights decaying (with a Gaussian distribution) with distance from the theoretical curve. The resulting moduli (*E* and *G*) quantified from the measured wavefields of nine (n=9) porcine cornea are shown in **Supplemental Figure 3**.

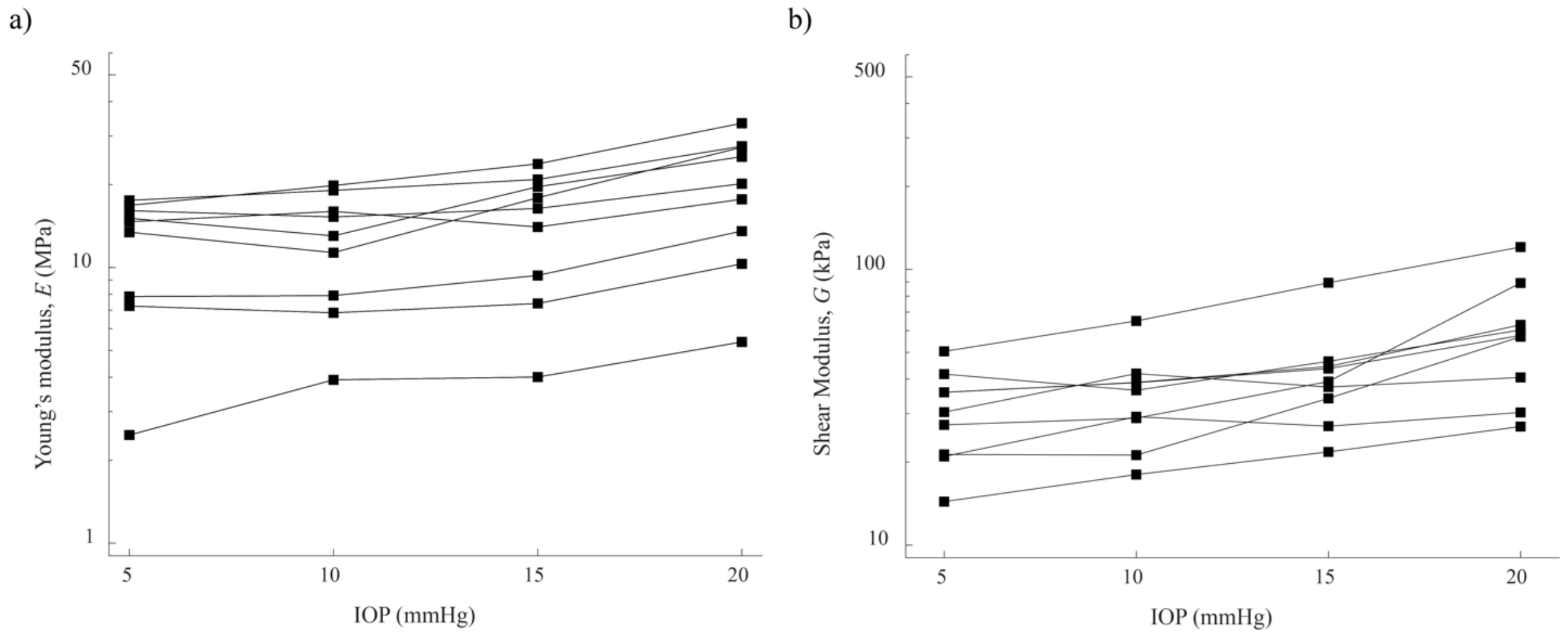

**Supplemental Figure 3.** (a) Young's modulus ($E$=3$\mu$) and (b) transverse shear modulus ($G$) in porcine whole globe samples quantified using the NITI model applied to AμT-OCE surface vibrations. The line connects moduli of each sample scanned at an IOP of 5, 10, 15, and 20mmHg to show individual variations in pressure-dependent moduli.

## 4. Shear Rheometry in porcine cornea

### 4.1. Parallel plate rheometry measurements

After scanning with AμT-OCE, corneal buttons were prepared for rheometry where the anterior and posterior surfaces of the circular stromal buttons were clamped between a parallel compression plate specially designed to match the size of the corneal surface. The surface of the 3D-printed and mounted head bracket had a diameter of 12mm and was rough enough to avoid tissue slippage.

For purely rotational strains, the stress-strain relation in a NITI material reduces to:

$$\begin{bmatrix}\tau_{yz}\\ \tau_{xz}\\ 0\end{bmatrix} = \begin{bmatrix}G & & \\ & G & \\ & & \mu\end{bmatrix}\begin{bmatrix}\gamma_{yz}\\ \gamma_{xz}\\ \gamma_{xy}\end{bmatrix}, \tag{S3}$$

where shear stress and strain relations may be described by:

$$\tau_{xz} = G\gamma_{xz}, \tag{S4}$$

$$\tau_{yz} = G\gamma_{yz}. \tag{S5}$$

The transformation from Cartesian coordinates to cylindrical coordinates gives:

$$\tau_{\theta z} = G\gamma_{\theta z}. \tag{S6}$$

A relative torsional twisting about an axis perpendicular to the surfaces was applied to deform the tissue in a state of pure transverse shear strain. A resulting sinusoidal oscillatory shear strain was measured according to:

$$\gamma_{\theta z}(t) = \gamma_{\theta z_0} e^{i\omega t}, \tag{S7}$$

where $i$ denotes the imaginary number, $\omega$ the angular frequency, and $\gamma_{\theta z_0}$ the peak shear strain amplitude (.1%.). The applied shear stress ($\tau_{\theta z}$) is described by:

$$\tau_{\theta z}(t) = \tau_{\theta z_0} e^{(i(\omega t + \delta))} \tag{S8}$$

where $\tau_{\theta z_0}$ is the shear stress amplitude and $\delta$ is the phase shift angle between the applied stress and the corresponding shear strain. The complex dispersive shear modulus, $G^*(\omega) = G' + iG"$, is defined by storage ($G'$) and loss ($G"$) terms. The shear stress and strain may then be described by:

$$\tau_{\theta z}(t) = G^*(\omega)\gamma_{\theta z}(t) \tag{S9}$$

The complex modulus is determined by dividing the shear stress by the shear strain (often in the frequency domain) to obtain the in-phase (storage) response to strain and 90 degrees out-of-phase (loss) modulus.

In general, collagen-rich tissue demonstrate an approximately linear behavior below 1%,[12] thus 0.1% peak strain was applied to the tissue during the frequency sweep (0.1-16 Hz). Frequency-dependent storage (*G*') and loss (*G*'') moduli, corresponding to the real and imaginary parts of $G^*$, were determined in the corresponding n=9 porcine cornea, with statistical mean and standard deviations for each sample plotted in **Supplemental Figure 4**.

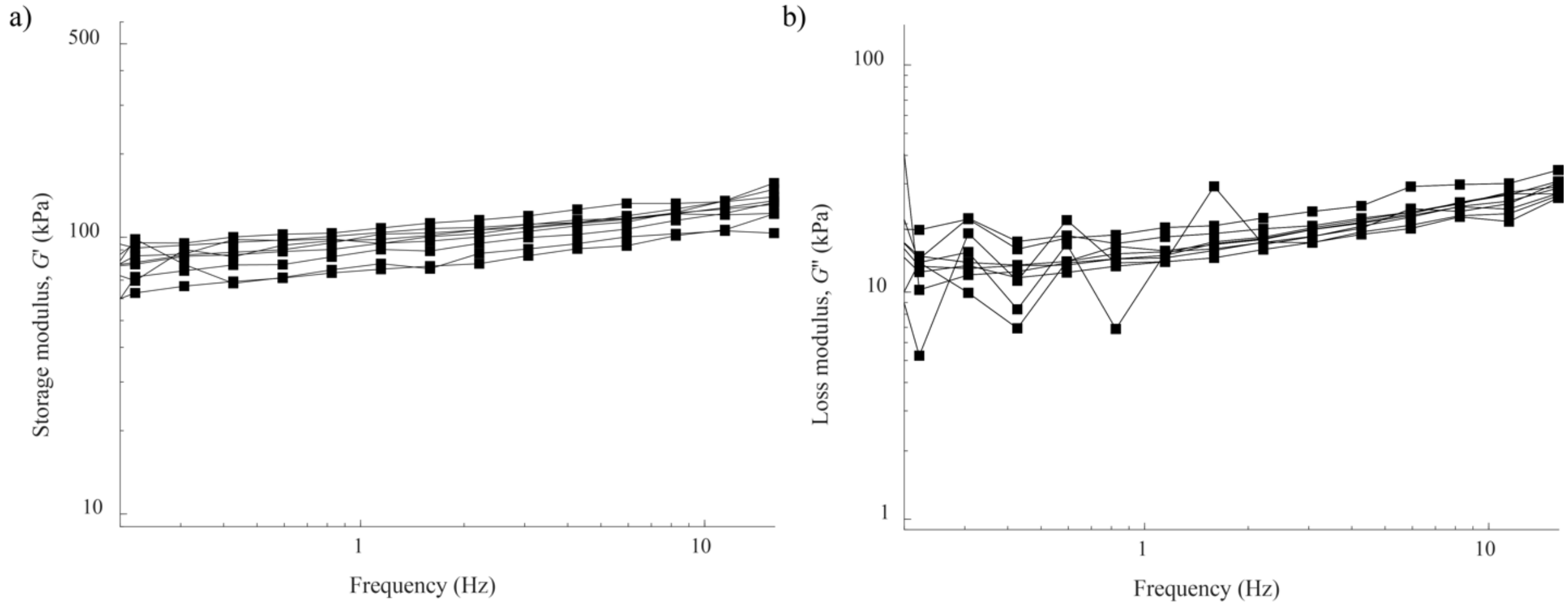

**Supplemental Figure 4.** Shear a) Storage (*G*') and b) Loss (*G*'') modulus measured during frequency-swept loading at .1% strain. The line connects moduli of each independent porcine cornea sample to show biological variations in frequency-dependent moduli.

4.2. Rheometry with and without sclera

Because transversally isotropic media require different loading geometries to measure both elastic moduli (*E* and *G*), excess sclera was left on the samples during rheometry to ensure sufficient surface area for clamping in tensile testing. Because a portion of the sclera was left on each corneal button during rheometry, the apparent stiffness due to excess tissue was tested in a set of three (3) porcine samples. Frequency-swept rheometry was perfomed on corneal buttons with a 2mm scleral boundary within 1 hour of porcine euthanasia. Immediately following two repeat frequency sweeps, the scleral ring was removed and only the cornea was exposed to a rotating shearing force for an additional two tests. The mean value of the two repeat tests (both with and without sclera) was used for each sample. The results are shown in **Supplemental Figure 5** and suggest that the excess tissue induced a 1.6-2.2 fold (depending on the frequency) increase in the shear modulus compared to rheometry of the cornea alone.

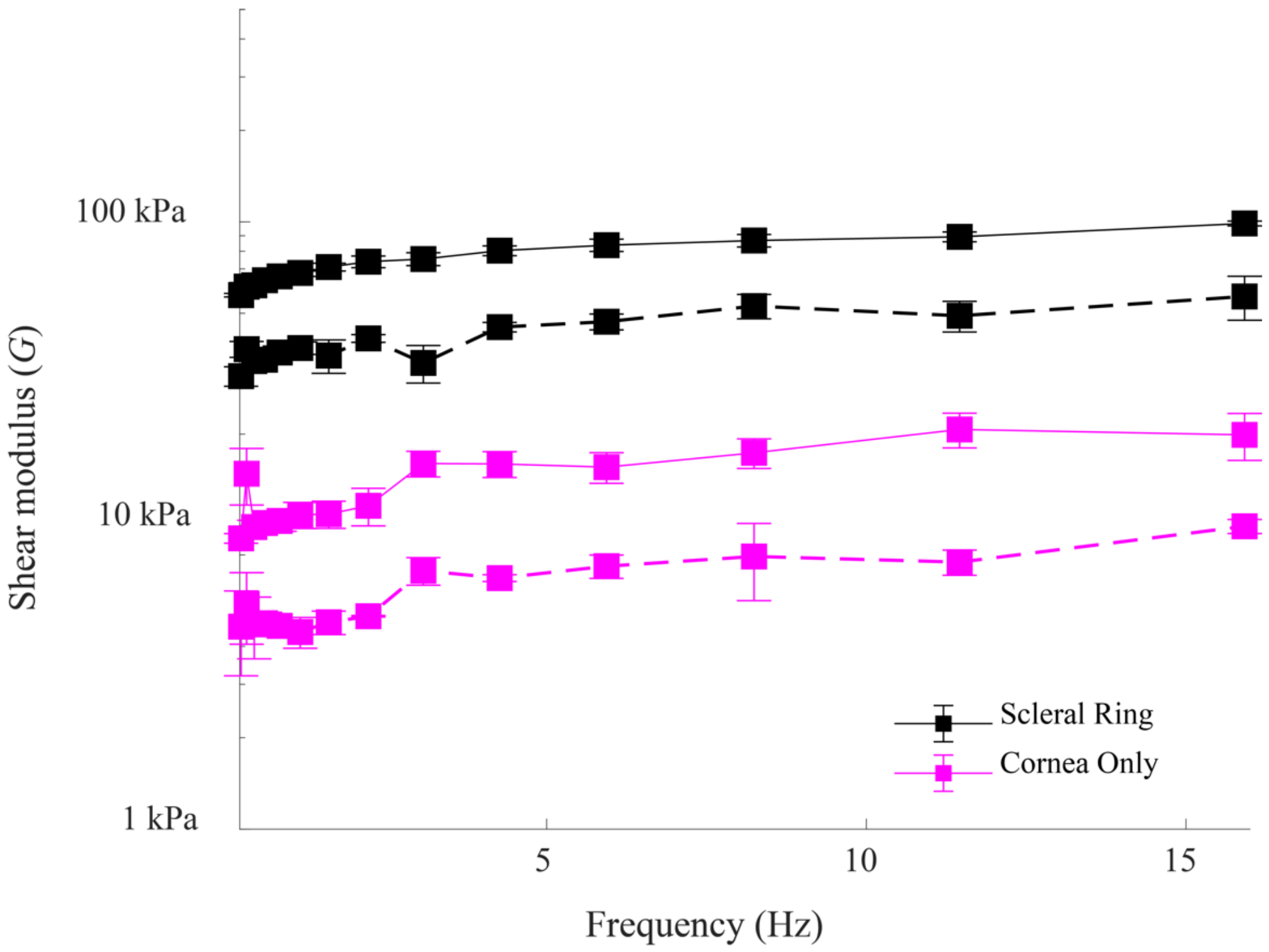

**Supplemental Figure 5.** Mean and standard deviation ($n$=3) of the storage (solid line) and loss (dotted line) modulus measured in samples with (black) and without (pink) the scleral ring attached. The excess sclera tissue increased the apparent modulus of the cornea.

## 5. Tensile Testing in porcine cornea

### 5.1. Extensometry measurements

Following rheometry, linear strips cut from corneal buttons were pneumatically clamped (2752-005 BioPuls submersible pneumatic grips, 250N max load) to the excess scleral tissue, leaving the cornea exposed between clamps. A 50mN pre-load was applied and the samples were stretched at 2mm/s (Instron model 5543) up to 10% strain. In this configuration, the tissue was subject to uniaxial tension along the $xx$-direction and the corresponding stress-strain relation described by:

$$\begin{bmatrix} \sigma_{xx} \\ 0 \\ 0 \end{bmatrix} = \begin{bmatrix} \lambda + 2\mu & \lambda & \lambda \\ \lambda & \lambda + 2\mu & \lambda \\ \lambda & \lambda & \lambda + 2\mu \end{bmatrix} \begin{bmatrix} \varepsilon_{xx} \\ \varepsilon_{yy} \\ \varepsilon_{zz} \end{bmatrix}. \tag{S10}$$

Defining the Young's modulus as $E = \sigma_{xx}/\varepsilon_{xx}$, the tissue response for the $xx$-loading direction is a function of both $\mu$ and $\lambda$:

$$E = \frac{\mu(3\lambda + 2\mu)}{\lambda + \mu}. \tag{S11}$$

Applying the incompressibility condition ($\lambda \rightarrow \infty$)**,** Eq. 11 reduces to $E = 3\mu$ for a given strain, demonstrating that tensile testing is sensitive to only the Young's modulus (independent of $G$).

The engineering stress ($\sigma$) in the corneal strips was calculated using the cross-sectional area of the central cornea,

$$\sigma = \frac{F}{wh}, \tag{S12}$$

where $w$ was the width, $h$ the thickness (as measured using a digital micrometer), and $F$ the force measured during elongation. The engineering strain ($\varepsilon$) was defined as

$$\varepsilon = \frac{\Delta l}{l_0} \tag{S13}$$

where $l_0$ was the initial length of the clamped cornea under a 50mN preload and $\Delta l$ the elongation values recorded by the extensometer. The stress-strain curve was determined for the corresponding n=9 porcine cornea (stripped in the temporal-nasal direction) by fitting a second order exponential to each set of raw data. (**Supplemental Figure 6a**). An additional second order exponential was fit to the data which included all (n=9) stress-strain values and was considered the representative curve for the n=9 porcine samples. The strain-dependent Young's modulus was defined by the instantaneous slope (Tangential Modulus) of the stress-strain curves (**Supplemental Figure 6b**).

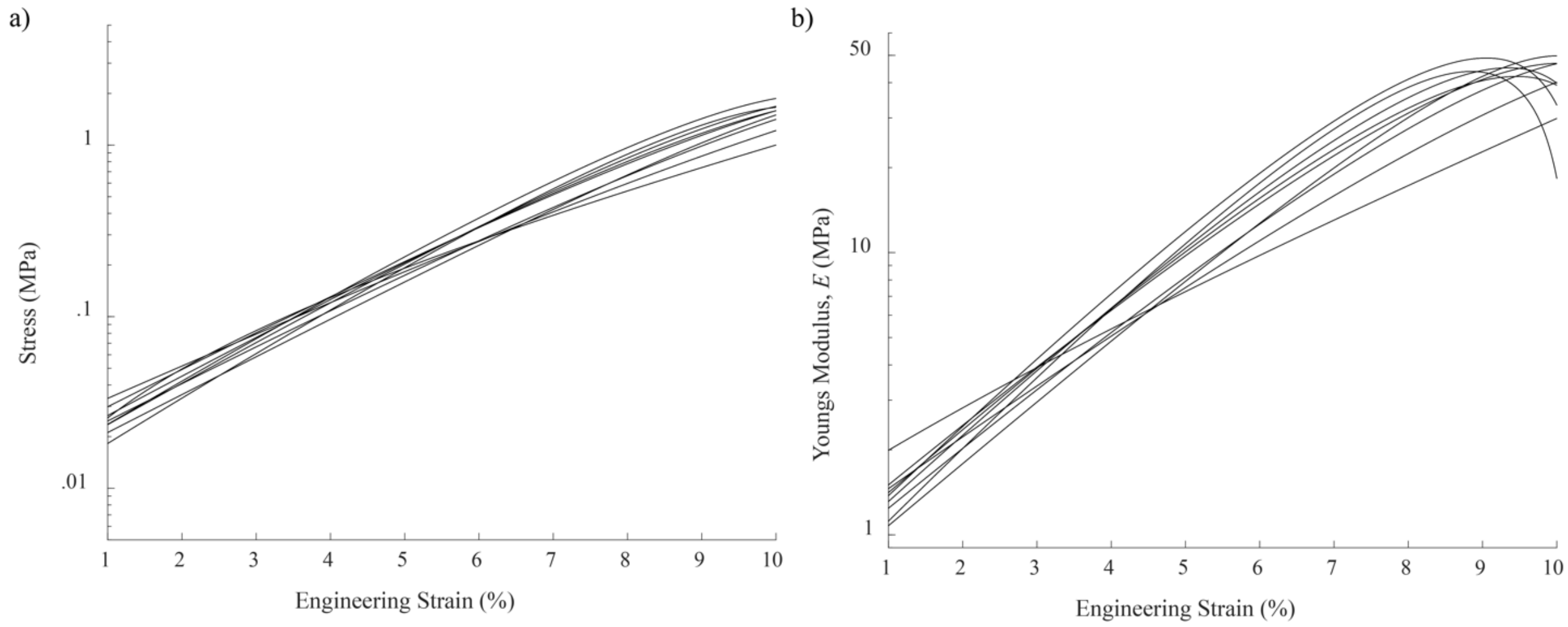

**Supplemental Figure 6.** a) Stress, Strain and b) Tangential modulus for 9 independent corneal strips. The displayed curve is the result of a second-degree exponential fit applied to the raw experimental data.

5.2. Compensation for Deformation State in Tension

It is well-documented that the cornea exhibits nonlinear moduli with strain. In a living organism, the cornea is highly pressurized by IOP, which produces an internal stress and preserves optimal corneal shape. However, when the cornea is extracted from the whole globe, its shape changes and internal stress is released. Therefore, it is very important to account for the cornea's deformation state and geometry change when comparing the modulus estimated by different methodologies. To compare OCE and tensile testing results, two factors producing different deformation states must be considered: (i) the cornea preserved its curvature in OCE measurements, while the curvature was flattened in tensile testing; (ii) the cornea was under biaxial tension due to the intraocular pressure inside the globe OCE measurements, while in tensile testing the cornea strip was loaded with a uniaxial stress.

Two models are presented below to address these factors. The first considers the geometry transition (curved to straight) as a reverse beam bending and derives an offset strain required to compensate modulus values estimated in the tensile test. The second uses the internal pressure as a common parameter to bridge uniaxial and biaxial loadings.

5.2.1. Reverse Bending Model

When the cornea strip is excised from a button, one can observe that its curved structure is generally retained. As the tensile force start to stretch the strip, it flattens to a straight piece under 50mN preload. This geometry transition introduces a complex pre-strain field in the corneal strip as shown in **Supplemental Figure 7a**, yielding part of the strip under compression and the rest under tension. This pre-strain may affect the accuracy of the tensile measurement since the part of the tissue under compression provides limited resistance against the force from the test device, leading to underestimation of the Young's modulus. As stretching continues during the test, we assume that underestimation stops when all compressional pre-strain is compensated and the whole strip is under tension. This happens when the strain applied by the tensile device equals the maximal compressional pre-strain $\varepsilon_{max}$. Here, we calculate $\varepsilon_{max}$ by reversing the beam bending model (as shown in **Supplemental Figure 7b**) since it is reciprocal to corneal stretching.

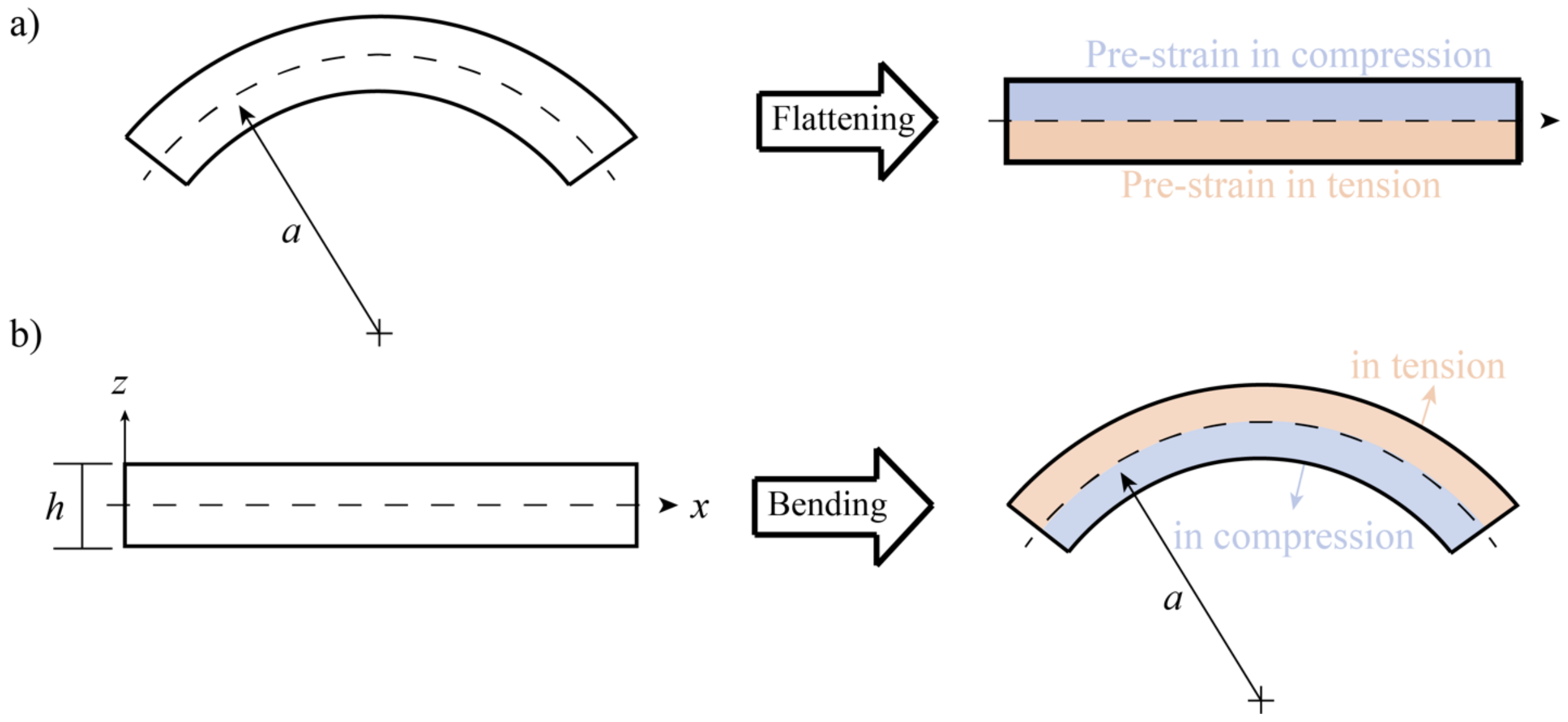

**Supplemental Figure 7.** a) Schematic showing the pre-strain distribution in the cornea strip when flattened at the beginning of tensile testing. b) Beam bending model for calculating the maximal compressional pre-strain in the cornea strip.

In **Supplemental Figure 7b**, the $x$-direction is along the longitudinal axis of the beam and located at the beam center, and the $z$-direction is along the thickness. The axial strain in the beam at different depths $\varepsilon_x(z)$ is defined as

$$\varepsilon_x(z) = \frac{z}{a}, \tag{S14}$$

where $a$ is the radius of curvature. The maximum compressional strain occurs at the bottom surface where $z = -\frac{h}{2}$, yielding

$$\varepsilon_{max} = -\frac{h}{2a}, \tag{S15}$$

In this notation $h$ is the thickness of the cornea. When the corneal strip is stretched by $\varepsilon_{max}$ during extension testing, the whole strip is assumed to be under tension and the tensile measurement has passed the underestimation region. Thus, we define the deformation state at this point to be equivalent to the "rest state" in the AμT-OCE measurements where there is no external pressure loading applied on the cornea (IOP = 0 mmHg). In other words, the rest state means the point where the geometric transformation from a spherical segment to the flat strip (or vise versa) is complete. Take the first cornea sample as an example, as shown in **Supplemental Figure 8a**, the solid circle at $\varepsilon = \varepsilon_{max} = 3.52\%$ and $E_{tensile}$ = 4.17 MPa marks the modulus estimated by tensile testing at the rest state.

5.2.2. Internal Pressure

AμT-OCE measurements were performed at different IOP (5, 10, 15, and 20 mmHg), which lead to different deformation states from the rest condition (0mmHg); therefore, this state change must be addressed when comparing Young's moduli estimated by both methodologies. Ideally, the comparison between methods should be performed at the same internal pressures; in AμT-OCE it is defined by IOP, and in tensile measurements the pressure is determined by the deformation level.

The internal pressure ($p_i$) is defined as one-third of the sum of all principal stresses:

$$p_i = \frac{1}{3}(\sigma_{11} + \sigma_{22} + \sigma_{33}). \tag{S16}$$

For the cornea in the AμT-OCE measurement, we may apply Timoshenko's thin shell theory with membrane approximation,[13] which yields the principal stresses

$$\sigma_{11} = \sigma_r = 0$$

$$\sigma_{22} = \sigma_\phi = \frac{P}{2}\frac{a}{h} \tag{S17}$$

$$\sigma_{33} = \sigma_\theta = \frac{P}{2}\frac{a}{h}$$

Where $P$ is the IOP, $a$ is cornea's radius of curvature, and $h$ is its thickness. The first principal stress $\sigma_r$ is 0 because AμT-OCE is performed on the traction-free boundary condition at the outer surface. Comparing to $\sigma_\phi$ and $\sigma_\theta$, $\sigma_r$ is much less significant because of the factor $a/h$ in $\sigma_\phi$ and $\sigma_\theta$, as $a$ is generally much larger than $h$ for the cornea. Therefore, we may simplify the calculation for deriving the internal pressure corresponding to different IOP in AμT-OCE measurements as

$$p_{i,OCE} = \frac{1}{3}\left(\frac{P}{2}\frac{a}{h} + \frac{P}{2}\frac{a}{h}\right) = \frac{P}{3}\left(\frac{a}{h}\right) \tag{S18}$$

Note that $p_{i,OCE}$ is geometry-dependent; that is, each cornea has a different value. For tensile testing, the internal pressure is given by

$$p_{i,tensile} = \frac{1}{3}\sigma_{11} \tag{S19}$$

where $\sigma_{11}$ is the uniaxial tensile stress.

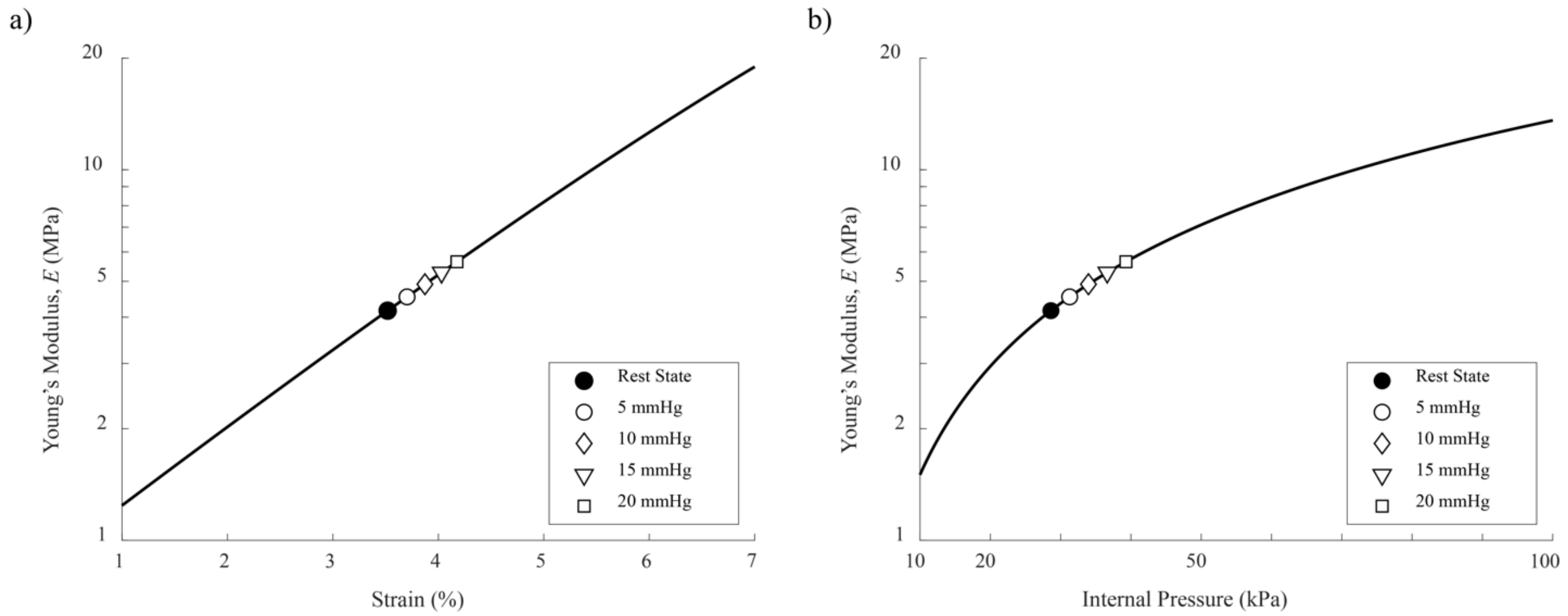

**Supplemental Figure 8:** Results of tensile testing from cornea sample No. 1. (a) Experimentally measured dependence of Young's modulus on strain in tensile test. The 'rest state' is determined by $\varepsilon_{max}$ calculated from the bending model. (b) Tensile modulus versus internal pressure. The internal pressure calculated with Timoshenko's thin shell theory corresponding to different IOP's is added to that of the rest state, yielding the effective IOP values for modulus estimates in tensile measurements.

**Supplemental Figure 8b** shows the same tensile testing results as that in **Supplemental Figure 8a**, but with estimated internal pressure on the abscissa. The internal pressure was calculated by Equation (S19). The rest state marked by the solid circle is determined by projecting the same Young's modulus $E_{\text{tensile}}$ = 4.17 MPa from **Supplemental Figure 8a**, which has the corresponding internal pressure of 28.61 kPa. From the rest state, internal pressure changes equivalent to those induced by IOP variations (5, 10, 15, and 20 mmHg) are added, leading to the internal pressures 31.88 kPa, 35.15 kPa, 38.42 kPa, and 41.69 kPa, and modulus estimations 4.62 MPa, 5.08 MPa, 5.53 MPa, and 5.97 MPa, respectively. Finally, the hollowed markers are projected back to **Supplemental Figure 8a** by referencing the same moduli from **Supplemental Figure 8b** to determine the uniaxial strains corresponding to the deformation states at the four IOP's. The process was performed for each cornea sample and the mean values reported in **Supplemental Table 1**.

**Supplemental Table 1** Estimated equivalent strain-dependent moduli

| **Inflation pressure**<br><br>**(OCE)** | **OCE Young's modulus, $E_{OCE}$**<br><br>**Mean ± SD (MPa)** | **Measured corresponding strain in tensile test** | **Calculated corresponding strain in tensile test**<br><br>**Mean ± (SD)** | **Tensile Young's modulus, $E_{tensile}$**<br><br>**Mean ± SD (MPa)** | **% Error**<br><br>$\frac{\lvert E_{tensile} - E_{OCE}\rvert}{E_{OCE}}$ |
|---|---|---|---|---|---|
| **5 mmHg** | 12.35 (±5.23) | **5.72%** | **3.95 (±.16)%** | 5.76 (±0.88) | 53% |
| **10 mmHg** | 12.57 (±5.53) | **5.78%** | **4.11 (±.14)%** | 6.19 (±0.90) | 51% |
| **15 mmHg** | 14.82 (±6.65) | **6.22%** | **4.25 (±.13)%** | 6.61 (±0.93) | 55% |
| **20 mmHg** | 20.05 (±9.16) | **7.10%** | **5.76 (±.88)%** | 7.03 (±0.96) | 65% |

**Supplemental References**